\documentclass[12pt,preprint]{aastex}
\usepackage{natbib}
\usepackage{float}

\def\be{\begin{equation}}
\def\ee{\end{equation}}
\def\bea{\begin{eqnarray}}
\def\eea{\end{eqnarray}}
\def\f{\frac}

\def\rm{\textrm}

\shorttitle{grb spectrum}
\shortauthors{zhao et al.}

\begin{document}
\title{Synchrotron spectrum of fast cooling electrons in GRBs}
\author{Xiao-Hong Zhao\altaffilmark{1,2,3},
Jin-Ming Bai\altaffilmark{1,2,3}}
\altaffiltext{1}{Yunnan Observatories, Chinese Academy of Sciences, 396 Yangfangwang, Guandu District, Kunming, 650216, P. R. China (e-mail:
zhaoxh@ynao.ac.cn)}
\altaffiltext{2}{Center for Astronomical Mega-Science, Chinese Academy of Sciences, 20A Datun Road, Chaoyang District, Beijing, 100012, P. R. China}
\altaffiltext{3}{Key Laboratory for the Structure and Evolution of Celestial Objects, Chinese Academy of Sciences, 396 Yangfangwang, Guandu District, Kunming, 650216, P. R. China}

\section{ABSTRACT} We discuss the synchrotron emission of fast cooling electrons in shocks. The fast cooling electrons behind the shocks can generate a position-dependent
inhomogeneous electron distribution if they have not enough time to mix homogeneously. This would lead to a very different synchrotron
spectrum in low frequency bands from that in the homogeneous case due to the synchrotron absorption. In this paper, we calculate
the synchrotron spectrum in this inhomogeneous case in a gamma-ray burst (GRB). Both the forward shock and the reverse shock are considered. We find
for the reverse shock dominated case, we would expect a ``reverse shock bump'' in the low frequency spectrum. The spectral bump is due to the combining
synchrotron absorption in both the forward and reverse shock regions. In the forward shock spectrum in the low frequencies
has two unconventional segments with spectral slopes of $\lesssim1$ and $11/8$. The slope of $11/8$ has been found by some authors, while the slope of $\lesssim1$ is new,
which is due to the approximately constant electron temperature in the optically thick region. In the future, simultaneous observations in multiple bands (especially
in the low frequency bands) in the GRB early afterglow or prompt emission phases will possibly reveal these spectral characteristics and enable us to identify the
reverse shock component and distinguish between the forward and reverse shock emissions. This also may be as a method to diagnose the electron distribution status
(homogeneous or inhomogeneous) after fast cooling in relativistic shock region.
\section{INTRODUCTION}
Gamma-ray bursts (GRBs) are the most powerful explosions in the universe. The standard fireball and internal-external shock models succeeded in explaining many
observations, especially the late GRB afterglow data. The internal shock model is one of the dominant models explaining the GRB prompt emission (Paczynski \& Xu 1994;
Rees \& M\'{e}sz\'{a}ros 1994), although it suffers several crucial
drawbacks (see Zhang \& Yan 2011 for a summary). The traditional internal shock model involves an unsteady relativistic wind (or multiple shells) driven by the GRB central engine.
A collision between two shells with different speeds would generate a forward shock and a reverse shock, in which the electrons are accelerated and produce the
prompt emission.

After a series of collisions, the merged shell runs into the circumburst medium and will also generate a forward shock
and a reverse shock (external shock) again. The forward shock (blast wave) model (Sari et al. 1998) is well consistent with the GRB afterglow data
(e.g. Galama et al. 1998). However, the expected bright reverse shock emission in the early afterglow is not observed in the majority of bursts (Roming et al. 2006).
Only a handful of bursts appear to be consistent with the reverse shock model (e.g. GRB 990123, Sari \& Piran 1999, however, see Meszaros \& Rees 1999 and Wei
2007, or recent bursts, GRBs 130427A and 160509A, Laskar et al. 2013, 2016).
This can be because the shells from the central engine are magnetized
and the reverse shock is not as strong as expected (e.g., Zhang \& Kobayashi 2005; Fan, Wei \& Wang 2004). The complexity of the reverse shock emission
(e.g. Kobayashi 2000 and Wu et al. 2003)
and its superposition with the forward shock emission also makes the identification of the reverse shock signature from the light curve difficult.

In GRB prompt and early afterglow phases, the magnetic field in the emission regions is strong enough that the energetic electrons accelerated in the shocks cool
by synchrotron or inverse Compton radiations within a much shorter time scale than the dynamic time\footnote{Here the dynamic time is the time in which the shock crosses
through the shell. },
which is called fast cooling. The spectrum in the fast cooling regime for a homogeneous electron distribution has been studied detailedly by some authors
(e.g., Sari et al. 1998).
The spectral slope below the synchrotron-self absorption (SSA) frequency is 2 or 5/2.
However, because electrons in the shocked shell are accelerated instantaneously and then cool before the shock crosses through
the shell, the electron with different (equivalent) temperatures can have not enough time to diffuse throughout the shell and thus
the electrons distribution is position-dependent, i.e, the electron distribution in the shell can be inhomogeneous.
The inhomogeneity will considerably affect the spectrum below the synchrotron absorption frequency. Granot et al. (2000)
have found the spectrum has an
unconventional segment with a slope of 11/8 in this case. We will show in the later sections that their result is for the forward shock.

As we know that the forward shock and the reverse shock would be produced in pairs. If the forward shock is advancing toward us, the reverse shock emission would cross
through the forward shock region before it reaches us. This can affect the reverse shock spectrum significantly if the forward shock is optically thick to the
reverse shock emission at low frequencies. Thus both shocks need to be considered when the emission from the two regions are considered.
Here we revisit the inhomogeneity problem and consider both the reverse shock and forward shock. We find some unprecedented spectral characteristics.
This paper is organized as follows. In section 3, we derive the electron
distribution in the inhomogeneous case. Section 4 is the calculation of the synchrotron emission. In the last section, we discuss the
possible application of our results.

\section{MODELS}
When a shock is crossing a shell, the electrons in a very thin layer (fluid element) behind the shock will be accelerated
instantaneously into a power-law distribution. With the shock
crossing the shell, more fluid elements become hot and cool down rapidly by synchrotron and inverse Compton (IC) radiation. These fluid elements have different
electron equivalent temperatures due to the electrons accelerated at different times. This would generate a temperature gradient: the electrons near the shock front
have not enough time to cool down but still remain hot while those far downstream would be cooler.
\subsection{ELECTRON DISTRIBUTION STRUCTURE}

\begin{figure}[!htb]
	\centering
	\includegraphics[width=3.2in]{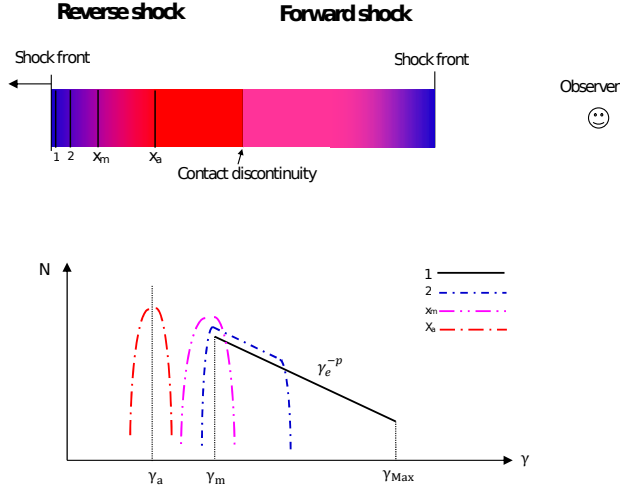}	
	\caption{Schematic of the equivalent temperature and the electron distribution in shocked shells. The color bar from blue to red in the upper part indicates the
		equivalent temperatures of the shells from hot to cool. Positions 1, 2 have different electron distributions due to different distances from the
		shock front and the corresponding electron distributions are shown in the lower part of this picture. $x_m$ is the position where all the electrons
                cool down to $\sim\gamma_m$.  $x_a$ is the position where the synchrotron peak
	frequency of electrons is equal to $\hat\nu_a$. See the text.} \label{fig1}
\end{figure}

Fig. \ref{fig1} is the schematic of electron status
in the forward and reverse shock regions. The equivalent temperatures in low frequency regions (the red region and the purple red regions in Fig. \ref{fig1}) in the two shocks
can be different due to different electron number densities.
We consider the time when the reverse shock right crosses through the shell and a photon is emitted from the reverse shock front. At this time,
the reverse shock emission reaches its peak.
With the traveling of the photon in the shocked shell toward the observer, more photons along the photon path in the shell will be emitted
and encounter absorption during their propagation.

We can define several critical positions behind shock front in each shocked region. The first is $x_m$. All the electrons within the fluid element at $x_m$
cool down to $\sim\gamma_m$ due to the radiation loss. The radiation power is $P_{r}=\f{4}{3}\sigma_Tc(\gamma_e^2-1)U_B(1+Y)$
including the cyclo-synchrotron and IC radiations, where Y is the Compton Y factor and $U_B=B^2/8\pi$ is the comoving magnetic field energy density.
$c$ and $\sigma_T$ are the light speed and the Thomson cross section, respectively. Below $x_m$, the electrons
at the fluid elements are hot and approximately remain a power law distribution with a high energy cutoff at some energy of $\hat\gamma_c$, while beyond it, most electrons
begin to cool (below $\gamma_m$) and approximately have a monoenergetic distribution with an energy of
$\hat\gamma_c$. Here $\hat\gamma_c$ is the energy to which the electrons cool down since the shock front passes some given position $x$ behind the shock front:

\bea
\hat\gamma_c(x)=\f{1+e^{-2bt(x)}}{1-e^{-2bt(x)}}
\eea
where $b=4\sigma_TU_{B}(1+Y)/3m_ec$ is a factor and $m_e$ is the electron mass. $t$ is the elapsed time since the electrons at x
are accelerated. Here the form of $\hat\gamma_c$ can apply to both the ultrarelativistic and non-relativistic
cases. When $2bt\ll1$, which corresponds to the ultra-relativistic case, we go back to the familiar form $\hat\gamma_c=6\pi
m_ec/\sigma_TB^2(1+Y)t$. The time $t$ can be given by
\bea
t(x)=\left\{\begin{array}{ll}
	4x/c&  \rm{reverse shock region} \\
	(6\Delta-2x)/c &  \rm{forward shock region}.
\end{array} \right.
\eea
We take into account the relativistic shock speed of $c/3$ in the shocked shell frame. In GRB, the internal shock is mildly relativistic for typical parameters
and thus the shock speed in the shocked shell would be $\lesssim c/3$, which may somewhat affect the final result. $\Delta$ is the comoving width of the shell.

The Compton Y factor is defined by the ratio of synchrotron photon energy density, including the contributions from the reverse shock region ($U_{syn,rs}$) and
the forward shock region ($U_{syn,fs}$), to the magnetic field energy density, i.e., $Y\equiv (U_{syn,rs}+U_{syn,fs})/U_{B}=[-1+\sqrt{1+4(U_{e,rs}+U_{e,fs})/U_B}]/2$
(Sari \& Esin 2001). $U_{e,rs}$ and $U_{e,fs}$ are the electron energy densities in the
two shocked regions. The corresponding electron number densities are $n_{0,rs}$ and $n_{0,fs}$. We neglect the second IC scattering, since it should
occur in the Klein-Nishina limit. Note that in the shock case, though the electron distribution can be inhomogeneous due to insufficient
diffusion time, the photon energes are approximately homogeneously distributed in the shocked shell because of the transparency of the shell to the spectral peak energy
($\nu_m$, synchrotron peak frequency corresponding to $\gamma_m$).
Thus we have $(U_{e,rs}+U_{e,fs})/U_{B}\simeq [(p-1)/(p-2)]m_ec^2(n_{0,rs}+n_{0,fs})\gamma_m/U_B$. Here we assume the heated electrons in the shocked regions have
the distribution of $dN/d\gamma_e\propto\gamma_e^{-p}$ and $p$ is the electron power-law slope.

The second critical position is $x_a$. As mentioned above, in the cool part of the shocked shell, the
electrons in each fluid element have approximately mono energy distributions. There is a position where the synchrotron peak frequency of electrons is equal to a
frequency $\hat\nu_a$ at which the optical depth is equal to 1.
$x_a$ can be obtained by solving
\bea \label{SSAfre}
\int_{0}^{x_a}\alpha_\nu(x,\hat\nu_a) dx=1,
\eea
where $\alpha_\nu$ is the absorption coefficient and will be given in the next section, $\hat\nu_a=0.45q_e B\hat\gamma_a^2/2\pi m_ec$ is used and
$\hat\gamma_a$ is the electron energy corresponding to $\hat\nu_a$.
Over $x_a$, the electrons begin to be thermalized with the peak of $\hat\gamma_a$ and the peak energy approximately remains a constant. An alternative derivation of $\hat\nu_a$
and $\hat\gamma_a$ is by balancing the cooling and SSA heating,
which should give similar results for $p<3$ (Ghisellini \& Svensson 1989).
We calculate $x_{a,rs}$ ($\hat\gamma_{a,rs}$) and $x_{a,fs}$ ($\hat\gamma_{a,fs}$) independently in the reverse and forward shock regions, respectively,
using the assumption that the radiation in one region does not affect the electron equivalent temperature in the other region.

We now can give the approximate electron distribution at each fluid element in the shocked shell:
\bea
\f{dN(x)}{d\gamma_e}\simeq\left\{\begin{array}{ll}
	\frac{n_0(p-1)}{\gamma_m}(\frac{\gamma_e}{\gamma_m})^{-p}(1-b\gamma_et)^{p-2} & \gamma_m<\gamma_e<\hat\gamma_c(x)~ ~ \textrm{or} ~ ~0<x<x_m\\
	n_0\delta[\gamma_e-\hat\gamma_c(x)]   & \hat\gamma_a<\gamma_e\leq\gamma_m ~ ~ ~ ~ ~~ \textrm{or} ~ ~~ ~x_m\leq x\leq x_a\\
	n_0\delta[\gamma_e-\hat\gamma_a] & \gamma_e\leq\hat\gamma_a ~ ~ ~ ~ ~ ~ ~ ~ ~ ~ ~ ~ ~\textrm{or}~ ~ x_a<x\leq \Delta,
\end{array}\right.
\eea
where $n_0$ is the electron number density. The first equation describes the evolution of the instantaneously injected electrons
with a power-law electron distribution (Kardashev 1962). The electron distribution in the region larger than $x_m$ in the shell are
described by a $\delta$-function. In the region larger than $x_a$, the
electron cooling by radiation is ineffective due to the balance between the SSA heating and the synchrotron+IC cooling and thus the electron peak energy roughly
remains a constant until the electrons cool by adiabatic expansion. Here we neglected the escape of electrons, since the
electrons should be confined by the magnetic field within a dynamic time scale.

\subsection{SYNCHROTRON SPECTRUM}
Using the above electron distribution, we can derive the synchrotron intensity $I_\nu$. The emission and absorption coefficients at each fluid element can be given by

\bea
j_\nu&=& \left\{\begin{array}{ll}\frac{1}{4\pi}\int d\gamma_eNP_\nu  & \hat\gamma_c\ge\gamma_m\\
	\frac{N_0}{4\pi}P_\nu(\hat\gamma_c)   & 2<\hat\gamma_c<\gamma_m \\
	\f{N_0}{4\pi}P_{\nu,cyc}& \hat\gamma_c<2
\end{array} \right. \\
\alpha_\nu&=& \left\{\begin{array}{ll}-\frac{1}{8\pi m_e\nu^2}\int d\gamma_e\gamma_ep_eP_\nu \frac{\partial}{\partial \gamma_e}\Big(\frac{N}{\gamma_ep_e}\Big)
	& \hat\gamma_c\ge\gamma_m\\
	\frac{N_0}{8\pi m_e\nu^2\hat\gamma_c^2}[\frac{d}{d\gamma_e}(\gamma_e^2P_\nu)]_{\gamma_e=\hat\gamma_c}
 & 2<\hat\gamma_c<\gamma_m \\
 \frac{N_0}{8\pi m_e\nu^2\hat\gamma_cp_c}[\frac{d}{d\gamma_e}(\gamma_ep_eP_{\nu,cyc})]_{\gamma_e=\hat\gamma_c}
 & \hat\gamma_c\leq2,
 \end{array} \right. \label{av}
\eea
where $P_\nu$ is the synchrotron spectral power and $P_{\nu,cyc}=\f{4\sigma_Tc}{3\pi}\f{p_c^2U_{B}}{\nu_L}\frac{2}{1+3p_c^2}e^{-\frac{2(1-\nu/\nu_L)}{1+3p_c^2}}$ is the
approximate cyclo-synchrotron power spectrum and
applies to electron energy $<2$ (Ghisellini et al. 1998). $p_c=\sqrt{\hat\gamma_c^2-1}$ and $p_e=\sqrt{\gamma_e^2-1}$ are the electron momentums, and $\nu_L$ is
the Larmor frequency. In the above second formula, we can find the absorption coefficient decreases with a power law of $\nu^{-5/3}$ below the synchrotron peak frequency
$\nu(\hat\gamma_c)$, while it exponentially decreases over it. Thus the synchrotron absorption optical depth in a region with a given electron energy
considerably decreases with the photon frequency increase when the photon frequency is higher than the synchrotron peak frequency the electron emits.

We can give the radiation intensity based on the radiative transfer equation:
\bea
I_\nu=\int_{0}^{\Delta} j_{\nu,rs} e^{-\int^{3\Delta}_{x}\alpha_{\nu,rs} ds}dx+\int_{\Delta}^{3\Delta}j_{\nu,fs}
	e^{-\int_{x}^{3\Delta}\alpha_{\nu,fs} ds}dx
\label{intensity}
 \eea

Here the upper limit of $\Delta$ corresponds to the peak time of the reverse shock emission in the observer frame.
The subscripts $fs$ and $rs$ denote the forward shock and the reverse shock, respectively.
The right terms of the equation are composed of two parts including
the contribution of the reverse and the forward shock regions. We take into account the fact that the reverse shock emission would cross through both
the reverse and forward shock regions before it reaches the observer. The upper limit of $3\Delta$ is due to the fact that when the reverse shock emission
catches up with the forward shock front, it travels $3\Delta$ in the shocked shell (or medium) frame.
We focus on the \emph{relativistic} reverse shock in the paper, which has stronger emission compared with the
Newton reverse shock, and we can neglect the shell spreading.

\section{RESULTS}
The spectral calculation depends on many unknown shock parameters. The parameters in GRB and its afterglow can be in a wide range. We first calculate the
spectra by adopting parameters in wide range. Then we take a group of plausible parameters in afterglow and calculate the spectrum.
\subsection{GENERAL PARAMETERS}

\begin{figure} [!htb]
	\centering{
		\includegraphics[width=4.2in]{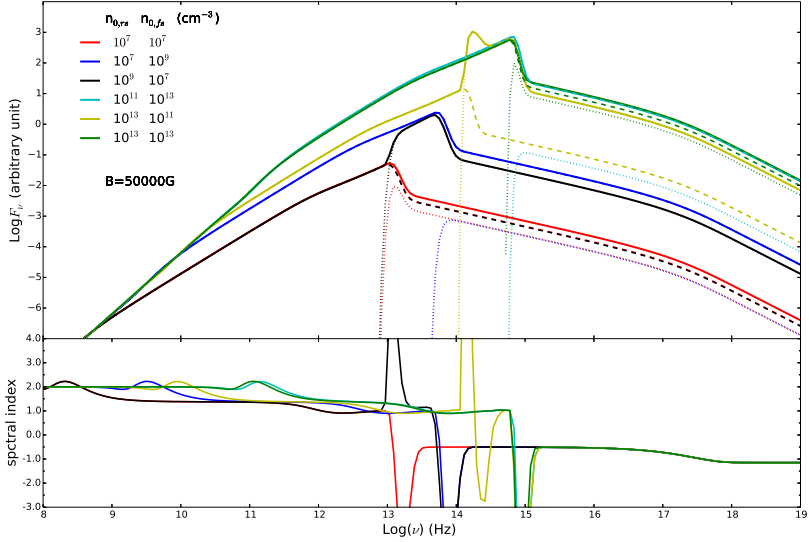}}	
\caption{The spectra (the upper panel) and the corresponding spectral index (lower panel) with the magnetic field
	of $B=5\times 10^{4}$ G. Different colors represent different parameters, marked in the figure. Other parameters are
	$\gamma_m$=$10^3$,$\gamma_{Max}$=$10^7$, p=2.3 and $\Delta$=$10^{12}$cm.
	In the upper panels, the solid lines represent the spectra from the shocked regions.
	The dotted lines and the dashed lines represent the contributions from the reverse shock and the forward shock, respectively.
The spectral slopes are shown in the lower panel.  } \label{fig2}
\end{figure}

\begin{figure} [!htb]
	\centering{
		\includegraphics[width=4.2in]{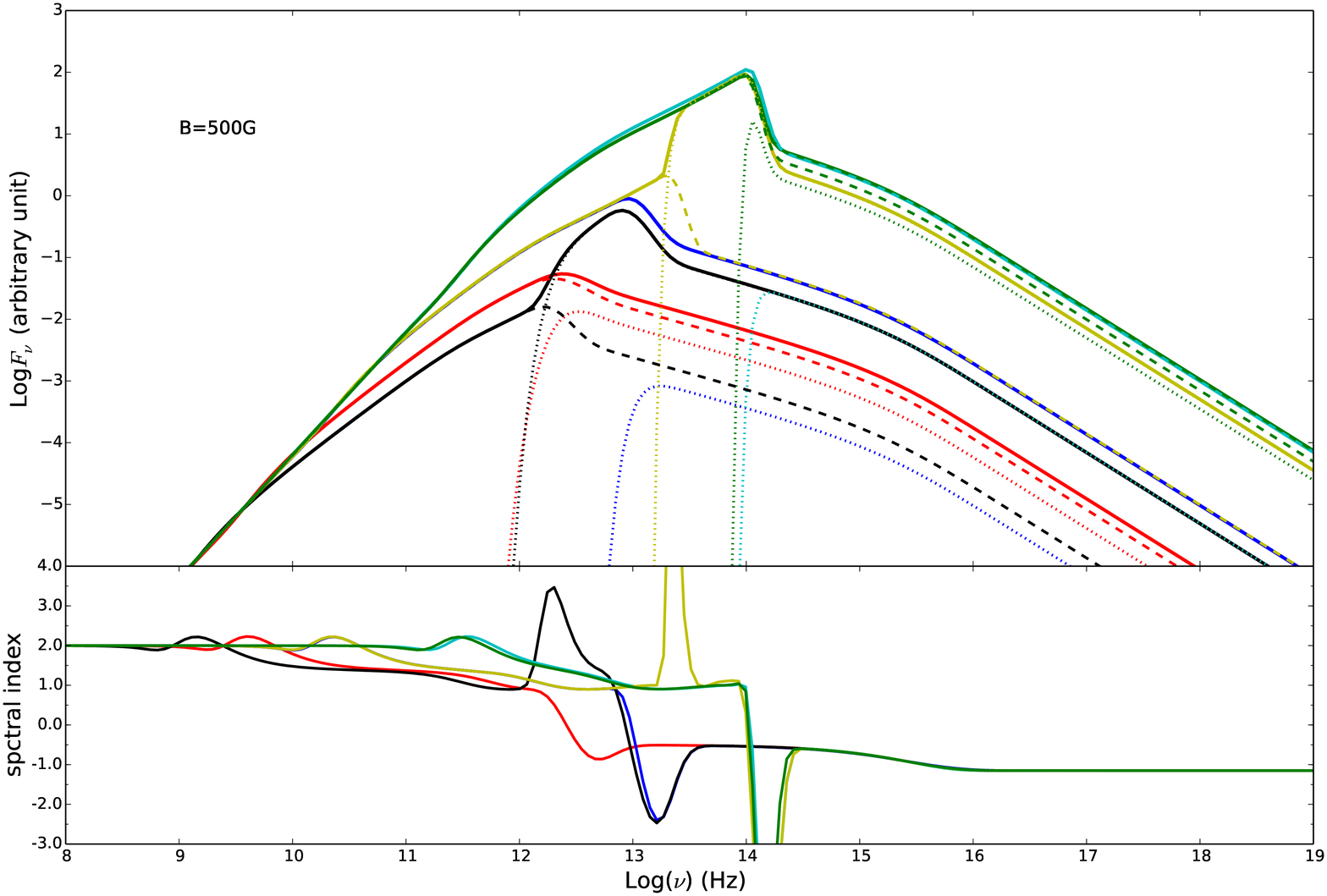}	}
	\caption{Same as in Fig. \ref{fig2} but for the magnetic field of $B=500$ G.} \label{fig3}
\end{figure}

\begin{figure} [!htb]
	\centering{	
		\includegraphics[width=4.2in]{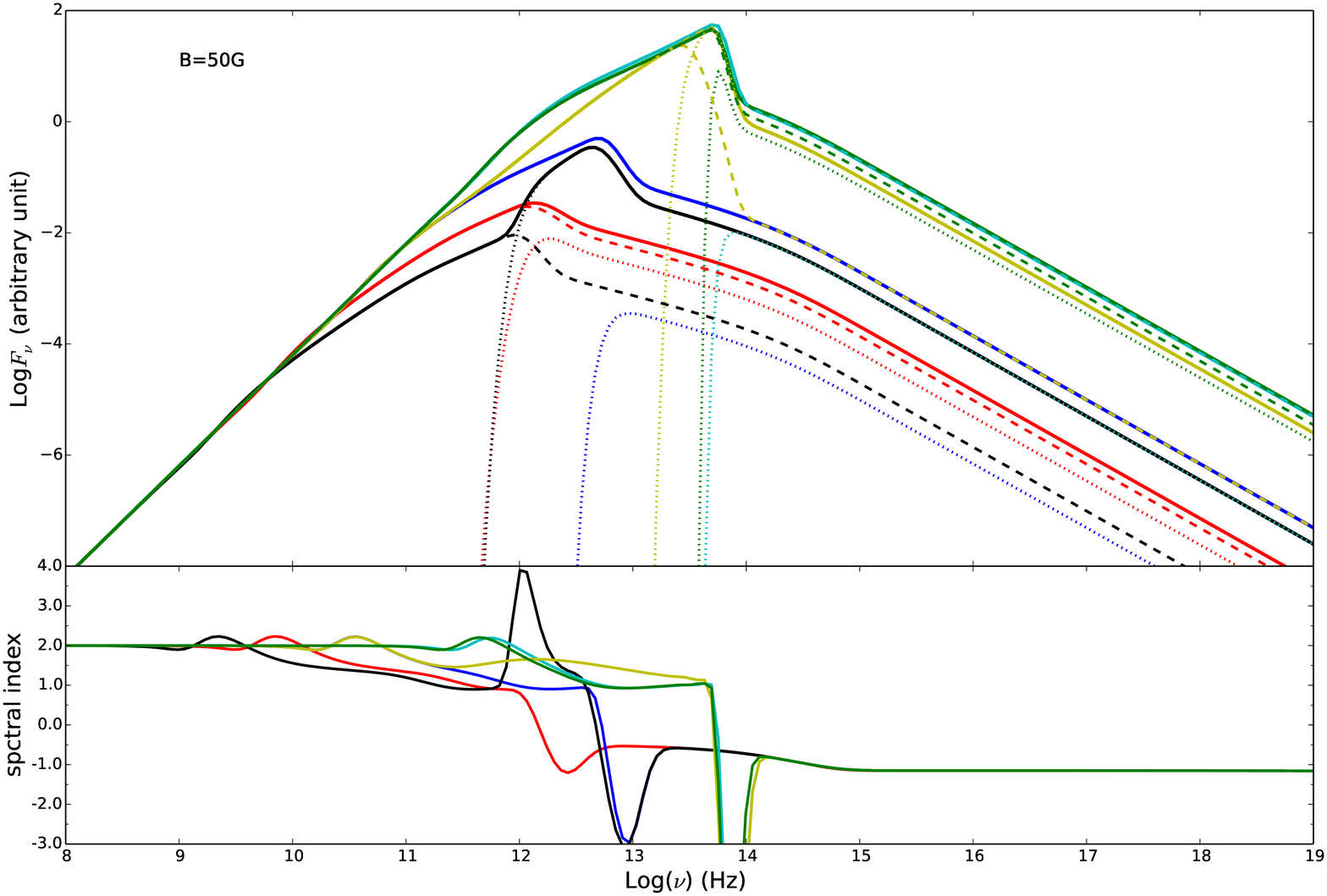}}
	\caption{Same as in Fig. \ref{fig2} but for the magnetic field of $B=50$ G} and shell width of $\Delta=10^{13}$cm. \label{fig4}
\end{figure}

Fig. \ref{fig2}-\ref{fig4} indicate the resulting spectra and the spectral slopes (in the comoving frame of the shocked shell).
One can find that the reverse shock
spectrum has a bump when the electron number density in the reverse shock region is much larger than that in the forward
shock region. This is because the photons from the reverse shock region encounter the absorptions in both the forward and reverse shock
regions. The synchrotron absorption from the forward shock region would lead to a sharp cutoff toward low frequencies due to the exponential
increase of the optical depth with frequency decrease. In low frequencies below the cutoff, the spectrum would be dominated by the forward shock emission.
With the increase of the photon frequency, the forward shock region gradually becomes optical thin to the reverse shock emission. The reverse
shock emission will generate a bump due to the piling-up of electrons at around $\hat\gamma_{a,rs}$. When the photon frequency is high enough that
the whole shell is transparent, the spectrum goes back to the optical-thin case.

When the electron density of the forward shock region is much larger than that in the reverse shock region, the forward shock emission is dominated.
The forward shock spectrum is quite different compared with the conventional homogeneous case. There is an unconventional
spectra segment in the low frequency bands, with a slope of $11/8$, which has been found by Granot et al. (2000). This segment results from the equivalent
temperature gradient in the shocked shell. The lower the photon frequency, the more approaching the forward shock front the emitted position. When the emitted position is
close enough to the shock front, electrons would be uncooled (with the average energy of $\sim\gamma_m$) and thus the spectral slope would be 2 due to the absorption
mainly from the electrons of $\gamma_m$. Another segment has a spectral index of $\sim1$, which is a new segment. This segment is due to the fact that with the
frequency increase, the electrons with the dominated contribution to the emission come to the SSA region (larger than $x_{a,fs}$) and the electron number approximately
linearly increase with the frequency, which leads to a spectrum with a slope of $\sim1$. It is worth noting that the segment of $\sim1$ also appears in the
reverse shock spectrum around the bump, which is due to the contributions of electrons in the reverse shock SSA region.

\subsection{APPLICATION TO THE EARLY AFTERGLOW}
\begin{figure} [!htb]
	\centering{	
		\includegraphics[width=4.2in]{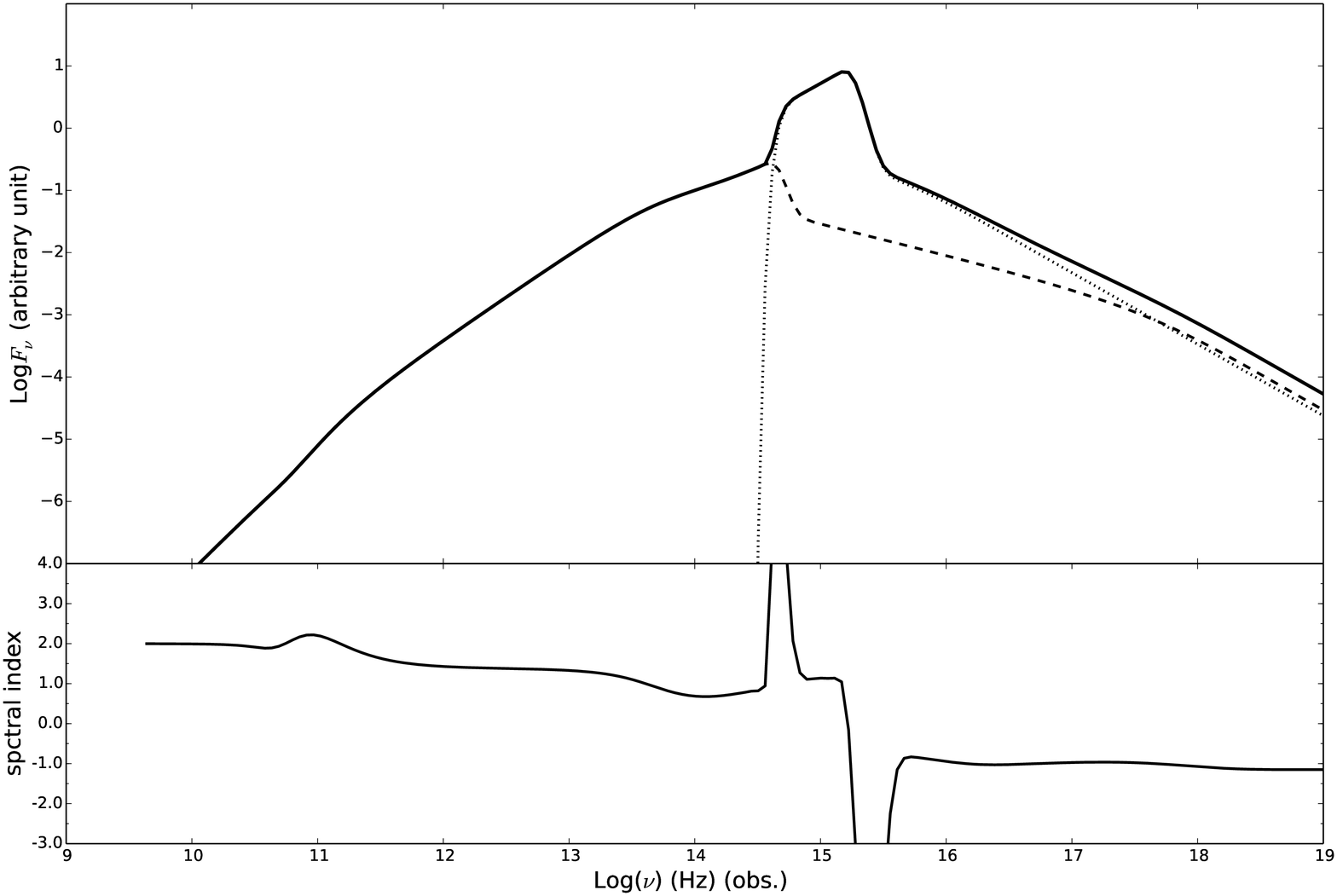}}
		\caption{The synchrotron spectrum for typical parameters at early afterglow phase in the wind environment. We use the frequency in the observer
			frame taking into account the relativistic motion of the shocked material. The adopted parameters are E=10$^{52}$ erg, A=3${\times}
			10^{35}$cm$^{-1}$, $\Gamma$=300,
		$\Delta$=$10^{14}$cm, $p=2.3$, $\epsilon_B=0.01$ and $\epsilon_e=0.1$.  In the upper panel, the solid line represents the spectra from the shocked regions.
		The dotted line and the dashed line represent the contributions from the reverse shock and the forward shock, respectively. The spectral slope is
	        shown in the lower panel. } \label{case}
\end{figure}

We present an application of our model in early afterglow in this section. The circumburst environment in GRBs is usually believed to be stellar
wind (Chevalier \& Li 2000) or interstellar medium (ISM). Here we consider the wind case (Wu et al. 2003). Using a group of plausible parameters
for GRB afterglows, we calculate the spectrum when the reverse shock just crosses the shell.
The number densities in the reverse and forward shock regions are $n_{0,rs}=2.1\times10^{9}$cm$^{-3} (\f{A}{3\times10^{35}})^{5/4}E^{-1/4}_{52}
(\f{\Gamma}{300})^{3/4}\Delta^{-7/4}_{14}$ and $n_{0,fs}=6.9\times10^{7}$cm$^{-3}(\f{A}{3\times10^{35}})^{7/4}E^{-3/4}_{52}
(\f{\Gamma}{300})^{5/4}\Delta^{-5/4}_{14}$, respectively. Here the circumburst number density profile is $n=A r^{-2}$,
where $A=\dot M/4\pi m_pv_w=3\times10^{35}$cm$^{-1}(\dot M/10^{-5}M_{\odot}$yr$^{-1})(v_w/10^{3}$km s$^{-1})^{-1}$. $E$ and $\Gamma$ are the energy and the
initial Lorentz factor of the shell, respectively. We use the usual notation $Q=10^mQ_m$ throughout
the paper. Assuming the magnetic field energy fraction of the internal energy $\epsilon_B$ in the two
regions are the same, then the magnetic fields in the two regions are also the same, which are given by $B=\sqrt{8\pi e_{fs}\epsilon_B}=1883.1$ G$
(\f{A}{3\times10^{35}})^{3/4}E^{-1/4}_{52}(\f{\Gamma}{300})^{3/4}\Delta^{-3/4}_{14}\epsilon_{B,-2}^{1/2}$ and $e_{fs}=1.4\times10^{7}$ erg cm$^{-3}
(\f{A}{3\times10^{35}})^{3/2}E^{-1/2}_{52}(\f{\Gamma}{300})^{3/2}\Delta^{-3/2}_{14}$  is the internal energy density in the two shocked region.
The Lorentz factor of the shocked shell and medium is
$\gamma_{sh}=33.9(\f{A}{3\times10^{35}})^{-1/4}E^{1/4}_{52}(\f{\Gamma}{300})^{1/4}\Delta^{-1/4}_{14}$. The minimum Lorentz factors in the two regions are
$\gamma_{m,rs}=\epsilon_e(\bar\gamma_{rs}-1)[(p-2)/(p-1)]m_p/m_e=144.8$, where $\bar\gamma_{rs}=4.4(\f{A}{3\times10^{35}})^{1/4}E^{-1/4}_{52}(\f{\Gamma}{300})^{3/4}
\Delta^{1/4}_{14}$ is the Lorentz factor of shocked shell relative to the unshocked shell and $\epsilon_e=0.1$ is the electron energy fraction of the internal
energy, and
$\gamma_{m,fs}=\epsilon_e\gamma_{sh}[(p-2)/(p-1)]m_p/m_e=1392.6$. The resulting spectrum is shown in Fig.\ref{case}. We can find the reverse
shock flux begins to exceed the forward shock flux from around 4$\times 10^{14}$ Hz. At V band ($\sim 5.4\times 10^{14}$ Hz), the reverse shock flux is roughly one
order of magnitude higher than the forward shock flux, i.e., the reverse shock emission is brighter than the forward shock emission by $\sim2.5$ mag at V
band. The reverse shock spectral bump and the forward shock spectral segments of $11/8$ and $\sim1$ clearly appears. The bump lies between the UV and soft X-ray
bands and seemingly will not be straightforward to observe. However, here we use some canonical parameters. In reality, GRB parameter ranges can be wide so that
the reverse shock bump will also lie in a wide range. Thus the bumps for some bursts will possibly move to the observable bands, such as optical bands.

\section{DISCUSSION}
In this paper, we consider the case that the electron distributions in shocks in GRB and its early afterglow phases are inhomogeneous due to the fast cooling of the electrons.
We calculate the spectra in this case by
considering the radiative transfer. We find the spectrum in the optically thick part is quite different from that of the usually considered homogeneous case.
For the reverse shock dominated case, the spectrum has a bump due to the combining absorption of the reverse and forward shock regions.
For the forward shock dominated case, there is an unconventional slope of $11/8$ in the spectrum, which is consistent with Granot et al. (2000 or Granot \& Sari 2002). 
There is also a new spectral segment of $\sim1$ following the segment of 11/8 toward high frequencies. Thus the spectral
slopes of the forward shock emission for the fast cooling are (2,11/8,1,-1/2,-p/2) from low frequency to high frequency. We also present an application to the
GRB early afterglow phase, i.e., the early forward and reverse shock phase of the external shock, with typical parameters. We find the above spectral characteristics,
such as the reverse shock bump and the forward shock spectral segments
of $11/8$ and $\sim1$, are all present. Such spectrum characteristics will possibly be observed in the future. It should be noted that such spectrum shape also possibly
presents in other objects with similar physical conditions to GRBs, such as active galactic nuclei (AGNs).

Actually, the reverse shock signature of the external shock is not clearly confirmed to date from the data.
Even the widely believed reverse shock cases, such as optical flashes of GRB 990123 and GRB 041219a, are still under debate.
M\'{e}sz\'{a}ros \& Rees (1999) and Wei (2007) considered the applicability of the internal shock model. As shown in this paper, the reverse shock emission has
very distinct features and behaviors as a bump above the forward shock emission if it dominates over the forward emission, which is easily recognized.
Thus this will be a method to identify the reverse shock emission from the spectrum. In the future if we observe the continuous spectrum in wide bands,
especially in the low frequency bands at early afterglow phase, or even several uncontinuous points in the low frequency bands, we could find the reverse shock
emission. While if the slopes of $11/8$ and/or $\sim1$ are observed, the spectrum should be from the forward shock.

If such spectra are detected in the prompt emission phase, this suggests that the prompt emission should come from a shock, most possibly
the internal shock. The shock model can naturally generate such an inhomogenity we discuss in the paper.
This may be used to distinguish between the internal shock model and other dissipation models, such as the magnetic reconnection model (e.g., Zhang
\& Yan 2011) where the electron distribution is close to the homogeneous case, since the magnetic reconnection occurs randomly.

Such spectral observation in turn suggests the electron distributions in shocks are indeed inhomogeneous, which may also present us new insight into the diffusing
process of fast cooling electrons behind a relativistic shock. The diffusion process of electrons is not well understood in physics, depending on some unknown factors
such as magnetic field strengthen and structure in the emission region, which requires more MHD simulation of shock. On the contrary, if the spectrum is characterized by
the slopes like 2 and/or 5/2 in low frequencies, then this may suggest the electron distribution of the emission region is homogeneous. Thus the spectrum
observation will be a probe of the mixability of electron distribution in emission region.

In observations, the wide band detections at early times, thanks to the rapid localization of GRBs by swift, have given us unprecedented insight into
GRB early afterglow physics (e.g., Zhang et al. 2006), though the coverage of bands is not enough to give us the spectral information yet.
However, the follow-up projects, especially in the low energies, are
increasing. The GRB and its early afterglow spectral detection in wider bands than now is promising. SVOM (Paul et al. 2011; Wei et al. 2016), NGRG (Grossan et
al. 2014) and UFFO (Grossan et al. 2012; Park et al. 2013) will
allow simultaneous or rapid follow-up observations of GRB in wide bands. GWAC (the Ground Wide Angle Camera) will possibly observe the optical emission and GFTs (the Ground
follow-up Telescope) will observe the near infrared emission during, after or even before GRBs (Paul et al. 2011). Early observations in longer wavelengths, such as
millimeter band and submillimeter band can be operated by EVLA (the Expanded very large Array \footnote{http://www.aoc.nrao.edu/evla/}) or ALMA (the Atacama Large
Millimeter/submillimeter Array \footnote{http://www.almaobservatory.org}). SKA (the Square Kilometer Array, e.g., Carilli et al. 2003) and FAST (the Five-Hunfred Aperture
Spherical Radio Telescope, Nan et al. 2011) will possibly contribute to the radio
observation. All these instruments will provide us more spectral information at the early afterglow time or even during the GRB,
allowing us to diagnose the revers shock and the forward shock signatures.

\begin{acknowledgements}
We thank Z. Li and X. F. Wu for useful suggestions and comments. We also thank an anonymous
referee for valuable comments to improve the paper. We acknowledge partial support by the Chinese Natural Science Foundation
(No. 11203067, 11133006, 11433004), Yunnan Natural Science Foundation (2011FB115 and 2014FB188) (X.H.Z.) and the Key Research Program of the
CAS (Grant NO. KJZD-EW-M06) (J.M.B.).
\end{acknowledgements}

\end{document}